\documentstyle[preprint,aps]{revtex}
\catcode`@=11 
\def\binrel@#1{\setbox\z@\hbox{\thinmuskip0mu
 \medmuskip-1mu\thickmuskip\@ne mu$#1\m@th$}%
 \setbox\@ne\hbox{\thinmuskip0mu\medmuskip-1mu\thickmuskip
 \@ne mu${}#1{}\m@th$}%
 \setbox\tw@\hbox{\hskip\wd\@ne\hskip-\wd\z@}}
\def\binrel@@#1{\ifdim\wd2<\z@\mathbin{#1}\else\ifdim\wd\tw@>\z@
 \mathrel{#1}\else{#1}\fi\fi}
\def\pmb{\let\next\relax\ifmmode\def\next{\mathpalette\pmb@}\else
 \let\next\pmb@@\fi\next}
\def\pmb@@#1{\leavevmode\setbox\z@\hbox{#1}\kern-.025em\copy\z@\kern-\ 
wd\z@
 \kern-.05em\copy\z@\kern-\wd\z@\kern-.025em\raise.0433em\box\z@}
\newdimen\pmbraise@
\def\pmb@#1#2{\setbox\thr@@\hbox{$\m@th#1{#2}$}%
 \setbox4\hbox{$\m@th#1\mkern.7794mu$}\pmbraise@\wd4
 \divide\pmbraise@18
 \binrel@{#2}\binrel@@{\mkern-.45mu\copy\thr@@\kern-\wd\thr@@
 \mkern-.9mu\copy\thr@@\kern-\wd\thr@@\mkern-.45mu\raise\pmbraise@\box 
\thr@@}}
\catcode`@=12 

\def\lsim{\mathrel{\rlap{\lower4pt\hbox{\hskip1pt$\sim$}}
    \raise1pt\hbox{$<$}}}         
\def\gsim{\mathrel{\rlap{\lower4pt\hbox{\hskip1pt$\sim$}}
    \raise1pt\hbox{$>$}}}         

\def\overleftrightarrow#1{\vbox{\ialign{##\crcr
    $\leftrightarrow$\crcr
    \noalign{\kern 1pt\nointerlineskip}
    $\hfil\displaystyle{#1}\hfil$\crcr}}}
    \long\def\caption#1#2{{\setbox1=\hbox{#1\quad}\hbox{\copy1%
    \vtop{\advance\hsize by -\wd1 \noindent #2}}}}

\def\frac#1/#2 {{\textstyle {#1\over #2}}}
\def\b#1{{\bf {#1}}}
\def\dag{\dagger}

\def\vrp{\b r {\scriptstyle ^{\,\prime}}}
\def\vr{\b r\,}
\def\s0{\sigma_0}
\def\si{\sigma}
\def\br{<\!}
\def\ke{\!>}

\def\a{\alpha} 

\def\be{\beta} 

\def\D{\Delta}

\def\k{\kappa} 

\def\l{\lambda}

\def\O{\Omega}
\def\o{\omega}

\def\t{\theta}

\def\underrightarrow#1{\vtop{\ialign{##\crcr
     
$\hfil\displaystyle{#1}\hfil$\crcr\noalign{\kern0pt\nointerlineskip}
     \rightarrowfill\crcr\noalign{\kern0pt}}}}

\def\frac#1/#2 {{\textstyle {#1\over #2}}}
\def\inf{\infty}

\def\b {\bf }
\def\dag{\dagger}
\def\bsi{{\pmb\sigma}}
\def\si{\sigma}
\def\Si{\Sigma}
\def\bga{{\pmb\gamma}}
\def\vep{\pmb \epsilon}
\def\ep{\epsilon}
\def\ga{\gamma}
\def\be{\beta}
\def\bal {{\pmb\alpha}}
\def\k{\kappa}
\def\kp{{\kappa'}}
\def\kpp{{\kappa^{\prime\prime}}}
\def\lpp{l^{\prime\prime}}
\def\o{\omega}
\def\O{\Omega}

\def\s0{\sigma_0}
\def\si{\sigma}
\def\br{<\!}
\def\ke{\!>}
\def\al{\alpha}

\def\arrow2#1{\vbox{\ialign{##\crcr
    $\leftrightarrow$\crcr
    \noalign{\kern-1pt\nointerlineskip}
    $\hfil\displaystyle{#1}\hfil$\crcr}}}
\def\tensor#1{\vbox{\ialign{##\crcr
    $\leftrightarrow$\crcr
    \noalign{\kern-1pt\nointerlineskip}
    $\hfil\displaystyle{#1}\hfil$\crcr}}}

\def\lsim{\mathrel{\rlap{\lower4pt\hbox{\hskip1pt$\sim$}}
    \raise1pt\hbox{$<$}}}         
\def\gsim{\mathrel{\rlap{\lower4pt\hbox{\hskip1pt$\sim$}}
    \raise1pt\hbox{$>$}}}         

\def\l{{\lambda}}

\def\cy{{\cal {Y} }}

\def\bcY{{\pmb{\cal {Y}}}}
\def\cA{{\cal {A}}}
\def\cB{{\cal {B}}}
\def\cC{{\cal {C}}}

\def\vj {{\bf j\,}}
\def\vJ{{\bf J\,}}
\def\vA{{\bf A\,}}

\def\vrp{{\bf r^{\;'}}}
\def\vr{{\bf r\,}}

\def\be{\begin{equation}}
\def\ee{\end{equation}}

\newcommand{\eq}[1]{\begin{equation}\label{#1}}
\newcommand{\eqr}[1]{Eq.~(\ref{#1})}

\newcommand{\eqn}{\begin{equation}}
\newcommand{\eqd}{\begin{displaymath}}
\newcommand{\ed}{\end{displaymath}}
\newcommand{\eqa}{\begin{eqnarray}}
\newcommand{\ea}{\end{eqnarray}}
\newcommand{\eqaa}{\begin{equation}\begin{array}{l}}
\newcommand{\eaa}{\end{array}\end{equation}}

\begin{document}
\begin{titlepage}

\hfill Preprint DOE/ER/40427-12-N96

\begin{center}
{\large \bf THE CHROMO-DIELECTRIC SOLITON MODEL:\\
QUARK SELF ENERGY AND HADRON BAGS\footnote{This paper is based 
in part on the doctoral dissertation of P. Tang, 
University of Washington, 1993.}}\\[8mm]

L. Wilets$^{(1)}$, S. Hartmann$^{(2)}$, and P.
Tang$^{(1),(3)}$\\[5mm]

$^{(1)}$ Dept. of Physics, Box 351560, University of Washington,
Seattle, WA 98195--1560\footnote{E-Mail: Wilets@nuc2.phys.washington.edu}\\
$^{(2)}$ Sektion Physik der Uni M\"unchen, 
Theresienstr. 37, D--80333 M\"unchen, Germany\footnote{E-Mail: 
Stephan.Hartmann@physik.uni-muenchen.de}\\
$^{(3)}$ Permanent address: Dept. of Technical Physics,
Peking University, Beijing, China\footnote{E-Mail: 
Tang@ibm320h.phy.pku.edu.cn}
\end{center}

\begin{abstract}
The chromo-dielectric soliton model (CDM) is Lorentz- and
chirally-invariant.  It has been demonstrated to exhibit dynamical
chiral symmetry breaking and spatial confinement in the locally uniform 
approximation. We here study the full nonlocal quark self
energy in a color-dielectric medium modeled by a two parameter Fermi 
function. Here color confinement is manifest.  The self energy thus
obtained is used to calculate quark wave functions in the medium which, 
in turn, are used to calculate the nucleon and pion masses in the one
gluon exchange approximation.  The nucleon mass is fixed to its 
empirical value using scaling arguments;  the pion mass (for massless
current quarks) turns out to be small but non-zero, depending on
the model parameters.
\end{abstract}

PACS: 12.39, 11.30.R, 14.20.D, 14.80.M

\end{titlepage}


\section{INTRODUCTION}
\label{sec:intro}

The chromo-dielectric soliton model (CDM) \cite{fai} is a
Lorentz-
and chirally invariant low-energy effective field theory based on
quantum chromodynamics (QCD). In order to simulate gluon condensates
and other scalar structures (as, e.g., $q\bar q$ pairs) inside
hadrons the QCD lagrangian density is supplemented by a scalar field
$\si$ that mediates the gluons through a color-dielectric
function.

Following arguments first given by T.D. Lee, a suitably
modeled color-dielectric function $\k (\si )$ guarantees absolute
color confinement \cite{Wi}. The assumed potential of the
scalar field is quartic and has two minima, one at zero and a second, 
deeper minimum at a
finite value identified as the vacuum value $\si_v$. In the absence
of quarks, the normal state of the $\si$-field is at the vacuum
value. In the presence of quarks and gluons,
the $\si$-field finds a minimum in
the vicinity of zero; the quarks and gluons dig a hole in the vacuum. 
This is
the origin of confinement in the model.

The CDM differs from the original Friedberg-Lee (FL) nontopological
soliton model \cite{fl} in the essential feature that there is no
direct quark-sigma coupling term. Thus the model is chirally
invariant for massless quarks. Krein {\it et al.}
\cite{krein} showed that for a  locally uniform dielectric medium,
chiral symmetry is dynamically broken if the strong coupling constant
or the inverse of the color-dielectric function exceeds a critical  
value.
Consequently, the quarks acquire an effective (``constituent'') mass.
The Nambu-Goldstone boson corresponding to this symmetry breaking has
been identified with the pion \cite{NJL}.

While the locally uniform model demonstrated {\it spatial
confinement} and the emergence of the pion, it did not demonstrate
{\it color confinement}. Furthermore, it was shown that the range of
non-locality of the quark self-energy was of the order of the typical
hadronic length scale and hence large compared with the soliton
surface. Therefore, it was deemed essential to investigate the
nonlocal quark self-energy for a realistic {\it and} self-consistent
soliton.

This is the problem we address in the present paper. We first obtain
the linearized (Abelian) gluon propagator in an inhomogeneous
color-dielectric medium. Because of the Abelian approximation, the
calculation is analogous to a problem in
electrodynamics.  The Schwinger-Dyson equation for the
quark propagator is solved along the
imaginary energy axis in order to avoid mass poles on the real energy
axis. Quark wave functions are obtained by solving the Dirac equation
with the self-energy playing the role of a nonlocal scalar potential  
which
is analytically continued to the real energy axis.

The mutual interaction between quarks and -- in the case of mesons --
antiquarks in hadrons is treated in the one gluon exchange
approximation (OGE). Corrections due to
center of mass motion are taken into account approximately. Using scaling
arguments, we fix the nucleon mass to its empirical value and
calculate the pion mass as a function of phenomenological
parameters. In the case of massless current quarks, the pion mass
turns out to be small but non-zero. Since a vanishing pion mass is
demanded by Goldstone's theorem, the calculated pion mass can be
considered a test of the approximation schemes applied \cite{weise}.

The paper is organized as follows. Sec.
\ref{sec:model} introduces the basic features of the
chromo-dielectric soliton
model. After deriving the equations for the gluon-propagator in
an
inhomogeneous medium (Sec. \ref{sec:gluon}), Sec. \ref{sec:sde}
addresses the formulation of the appropriate Schwinger-Dyson
equation
for the quark self-energy. This self-energy is used in Sec.
\ref{sec:wave} as an effective nonlocal quark potential in order
to
determine quark wave functions in a bag. Sec.
\ref{sec:numerical}
contains details of the numerical solution of the corresponding
equations and presents results for the self energy.
Sec. \ref{sec:hadronic} describes the calculation of hadronic
properties in the OGE-approximation and, finally, Sec.
\ref{sec:summary} sums up the main results and discusses future
prospectives.


\section{THE MODEL}
\label{sec:model}

The CDM lagrangian density is given by \cite{krein}
\be
{\cal L}_{CDM} = \bar q( {\rm i} \gamma^\mu\partial_\mu
+ g_s\, {1\over 2}\,\lambda^a\;A^a_\mu\,\gamma^\mu - m_f) q
- \k(\si){1 \over 4}\, F^a_{\mu\nu}F^{a\mu\nu} + {1\over
2}(\partial_\mu \si)^2
-U(\si) + {\cal L'} \,,
\ee
\be
F^a_{\mu\nu} = \partial_\mu A^a_{\nu} - \partial_\nu A^a_{\mu}
+ g_s f^{abc} A^b_{\mu} A^c_{\nu} \,,
\ee
where the color $SU(3)$ structure constants satisfy
$\left[ \lambda^a , \lambda^b \right] = 2 i f^{abc}
\lambda^{c}$,
$q$ are the quark fields, $A^a_{\mu}$ are the gluon fields,
$\si$  is
the effective scalar field which determines the effective
color-dielectric function $1 \ge \k (\si ) \ge 0$, and ${\cal
L'}$
contains any necessary counter terms, gauge fixing term, or
ghosts. It
is evident that the model is gauge invariant. $m_f$ is the
quark
flavor (current-) mass matrix. Throughout this paper we will set
the
current quark masses equal to zero, so that the model is also
explicitly chirally invariant. The color-dielectric function $\k
(\si)$
mediates the gluon field and is designed to guarantee
color confinement. It has been shown \cite{Wi} that the
following
assumptions must  be satisfied: $\k(0) = 1, \k(\si_v) =
\k'(\si_v)
= \k'(0) = 0$. These constraints are satisfied, e.g., by
\be
\k(\si) = 1 + \t(x)\,x^n\,(n x-(n+1)) \,, \qquad  n >2,
\label{2kappa}
\ee
with $x = \si/ \si_v$.  The vacuum-value of the
$\si$-field is denoted by  $\si_v$.  We choose $n=3$ for 
simplicity so that $\k(\si)$ is continuous at $x=0$.

Analogous to the FL-model, the potential of the $\si$-field
is
given by the quartic form
\eq{usig}
U(\si) = {a\over 2!}\si^2 + {b\over 3!} \si^3 + {c\over 4!}
\si^4 +
B\,.
\ee
The ``bag constant'' $B$ is fixed in terms of the other
model-constants so that $U(\si_v) = 0$. In the FL-model,
$U(\si)$ is
chosen to be quartic in order to make the model renormalizable.
Although our model is not renormalizable (due to the presence of
$\k
(\si)$) we stick to this form in order to minimize the numbers
of
free parameters in the model. We identify $U''(\si_v) \equiv
m^2_{GB}$ with the lowest $0^{++}$ glueball mass and require
$U'(\si_v) = 0$ \cite{Wi}.

We now discuss the divergences of the model in more
detail. The model exhibits both infrared and ultraviolet
divergences. The origin of the {\it infrared} divergence is the
same
as for the MIT bag model. For a spherical bag, for example,
the electric monopole term of the quark self-energy diverges as
$r\rightarrow \inf$. This
happens in the CDM, if the color-dielectric constant vanishes
as
$r\to\infty.$  The
infrared divergence is thus associated with {\it color
confinement}.
However, for a color singlet bag, no infrared divergence
occurs since the self and mutual interaction terms cancel when
ladder diagrams for the mutual interaction are properly
calculated. The monopole term of the self energy is ignored
in the MIT bag model. Since this term is the source of color
confinement
in our model we cannot neglect it. In our calculations we choose a 
Fermi function shaped spherically symmetric color-dielectric function as 
displayed in figure 1.

The {\it ultraviolet} divergence is associated with the point
nature
of quarks. It is shown in ref. \cite{krein} that the effective
quark
mass, which is generated because of dynamical chiral symmetry
breaking, goes to infinity if the
color-dielectric function approaches zero. This divergence is
thus
connected with {\it spatial confinement}.  We handle this
divergence by introducing an energy cutoff (asymptotic freedom).
For
numerical reasons, we regulate the infrared divergence by
adjusting
$\kappa_v = \kappa(\si_v)$ to a small, non-zero value, and discuss the 
limit $\kappa_v \rightarrow 0$.


\section{THE GLUON PROPAGATOR}
\label{sec:gluon}

We assume that parts of the non-Abelian effects are effectively
included in the $\si$-field. This allows us to approximate the
gluon
field by its Abelian part. Hence the gluon field equations are
formally identical to Maxwell's equations in an inhomogeneous
medium
characterized by a time-independent color-dielectric function
$\k (\vr )$.
The field equations for the vector potential $A_\mu (\vr,\,t)$ 
read (we follow here references \cite{BGW} and \cite{TW}):
\eq{4max}
\partial^\mu \k (\vr)[\partial_\mu A_\nu - \partial_\nu A_\mu]
= J_\nu (\vr,\,t) \,.
\ee
Since the Abelian approximation destroys gauge invariance, the
choice
of gauge is part of the approximations. We choose the Coulomb
gauge
defined by
\be
\pmb{\nabla}\cdot (\k{\vA}) = 0\,.
\ee
The $\nu=0$ component of \eqr{4max} yields
\be	
-{\pmb\nabla} \k (\vr) \cdot {\pmb\nabla} A_0 (\vr,\,t) =
J_0(\vr,\, t)\,.	
\ee
The time-time component of the Green's function, $D^{00}$,
defined by
\eqn
A_0(\vr,\, t) = \int d^3\!r'D^{00}(\vr,\,\vr')J_0(\vr',\,t)\,,
\ee
satisfies the equation
\eq{4eg0}
- \pmb\nabla \k (\vr ) \cdot \pmb\nabla
D^{00}(\vr,\,\vr') = \delta^3(\vr-\vr')\,.				
\ee
Note that $D^{00} (\vr,\,\vr')$ is instantaneous. 


Now consider the $\nu = i$ components of \eqr{4max}
\eq{4vec}
\k \partial_t^2 {\vA} - \nabla^2 (\k {\vA}) + \pmb\nabla\times
\left( \k {\vA} \times \pmb\nabla\ln \k \right)
= {\vJ} -\k \pmb\nabla \partial_t  A_0\equiv {\vJ}_{tr}\,.
\ee
The transverse current defined by \eqr{4vec} can be expressed
in
terms of
$\b J$ using the time-time Green's function:
\eqn
{\vJ}_{tr} (\vr,t) = {\vJ}(\vr,t)-\k(\vr) \pmb\nabla\int
d^3\!r'D^{00}(\vr,\,\vr') \partial_t {J_0}(\vr',t)\,.
\ee
Using current conservation, $\partial_t  J_0 + \pmb\nabla\cdot
{\vJ} = 0$, and performing a partial integration, we obtain
\eq{42.8}
{\vJ}_{tr} (\vr,t) = {\vJ}(\vr,t) - \k (\vr) \pmb\nabla \int
d^3\!r'
\left(\pmb\nabla'D^{00}(\vr,\,\vr') \right) \cdot \vJ
(\vr',t)\,.		
\ee

We now Fourier transform the time dependence of $\vJ(\vr,\,t)$
and
$\vA(\vr,\,t)$ to $\vJ(\vr,\,\o )$ and $\vA(\vr,\,\o )$. The 
Green's
function corresponding to \eqr{4vec} satisfies
\eq{4tran}
-\left[\nabla^2 +\o ^2 + \pmb \nabla\times
\left(\pmb\nabla\ln \k \right)\times \right]
\k \tensor D(\vr,\,\vr') = \tensor  \delta_{tr}(\vr,\,\vr') \,,
\ee
where the components of the transverse delta function are given
by
\eq{2.11}
\delta^{ij}_{tr}(\vr,\,\vr') = \delta^{ij} \delta^3(\vr-\vr')
-\k (\vr ) \partial^i{\partial'}^j D^{00} (\vr,\,\vr')\,.
\ee

In this paper we will restrict ourselves to spherical bags. In
this
case the Green's functions can be decomposed in terms of
spherical
harmonics \cite{BGW}:
\eq{5D}
{D^i}_{i'} \to \tensor D(\vr,\vrp;\o) = \sum_{jll'm_l}
 d_{jll'}(r,r';\o)
{\overleftarrow Y_{jlm_l}}(\O) \overrightarrow Y_{jl'm_l}^*(\O')
\,,		
\ee
\eq{5D0}
D^{00}(\vr,\vrp) = \sum_{lm_l} d^0_{l}(r,r')
{Y_{l m_l}}(\O) Y_{l m_l}^*(\O') \,.
\ee
Some of the tensor components are shown in figs. 2 and 3. 
The $\k (r)$ parameters are again $R = 0.8 $ fm,
$A = 0.1 $ fm, and $\k_v = 0.1$.

It should be noted that the Green's functions do not carry any
color indices.  This results from the fact that the medium is
color-neutral so that  $D_{\mu \nu}$ has a trivial (diagonal)
color structure.

Details of the derivation and solution of equations (\ref{4eg0})
and (\ref{4tran}) are given in \cite{BGW}, an important
correction is reported in ref. \cite{TW}.


\section{THE SCHWINGER-DYSON EQUATION IN THE QUARK-GLUON SECTOR}
\label{sec:sde}

Being now in the possession of the gluon propagator in the
cavity we can study the Schwinger-Dyson equation for a single
quark in a cavity. In the course of this calculation we need not
refer to the $\si$-field.

The Schwinger-Dyson equation reads (in $(\o, \vr)$-space):
\eq{5sd}
\Si(\vr,\vrp;\o) = {\rm i} \al'\int_{-\inf}^{\inf}  d\o'
D_{\mu\nu}(\vr,\vrp;\o') \ga^\mu
S(\vr,\vrp;\o-\o') \ga^\nu \,,				
\ee
with $\al' = (4/3) g_s^2 /2\pi$. In Eq. (\ref{5sd}) we have
already approximated the one-particle irreducible quark-gluon
vertex $\Gamma^\mu$ by the bare one.

It is easy to show that both the gluon propagator $D_{\mu \nu}$
and the quark propagators $G$ do not have poles off the
real $\o$-axis. So according to the Schwinger-Dyson
equation, the self energy $\Si$ should have no pole of the
real $\o$-axis as well.
Thus we perform a Wick rotation, $\o \to {\rm i} y$ and study
the self energy first for imaginary $\o$. The ``rotated''
Schwinger-Dyson equation now reads:
\eq{5sdi}
\Si(\vr,\vrp; y) = -\al'\int_{-\inf}^{\inf} dy'
D_{\mu\nu}(\vr,\vrp;
y') \ga^\mu S(\vr,\vrp; y-y') \ga^\nu \,.
\ee
Simultaneously, the Dirac equation for the quark propagator has
to be satisfied:
\eq{5dirac}
(\o \ga^0 - \bga \cdot {\b p} - \Si) S = \delta^3 (\vr - \vrp) \,.	
\ee
To simplify the notation we have used the shorthand $\Si S$ for 
$\int
d^3 r_2 \Si(\vr,\vr_2;\o) S(\vr_2,\vrp;\o)$. Throughout this
paper repeated spatial coordinates are integrated over.

We now define the following hermitian functions,
\eq{5def1}
G = - S \beta,\; \; V = \beta \Si \,.
\ee
Eqs. (\ref{5dirac}) and (\ref{5sd}) then become
\eq{5dirac2}
(- \o + \bal \cdot {\b p} + V)G = \delta^3(\vr- \vrp) \,,
\ee
\eq{5sd2}
V(\vr,\vrp; y) = \al'\int_{-\inf}^{\inf} dy'
D_{\mu\nu}(\vr,\vrp; y')
\al^\mu G(\vr,\vrp; y-y') \al^\nu \,.
\ee
Here $\al^\mu \equiv (1,\pmb\al)$ is used for notational
convenience only. It is obviously not a Lorentz vector.

From the coupled Eqs. (\ref{5dirac2}, \ref{5sd2}), the
hermiticity of $G$ and $V$ can be verified:
\eq{5p1}
G^\dag(\vr,\vrp;\o) = G(\vr,\vrp;\o^*) \,,
\ee
\eq{5p2}
V^\dag(\vr,\vrp;\o) = V(\vr,\vrp;\o^*) \,.
\ee
The hermitian conjugation includes the interchange of the
arguments $\vr$ and $\vr'$:
\eq{5herm}
V_{ij}^\dag(\vr,\vrp;\o) \equiv V_{ji}(\vrp,\vr;\o)^* \,.
\ee


\subsection {Angular decomposition of the quark propagator}
\label{subsec:angular}

In order to solve the coupled Eqs. (\ref{5dirac2}, \ref{5sd2})
numerically, we make an angular decomposition of the
appropriate quantities. For spherically symmetric
color-dielectric functions $\k (r)$, the hermitian properties
$G$ and $V$ can be decomposed \cite{WK,Gy}
\eq{5defG}
G(\vr,\vrp;\o) =  \sum_\k
\left( \matrix{ g_\k^{11}(r,r';\o) \pi_{\k}
& g_\k^{12}(r,r';\o) i\si_r \pi_{-\k} 	\cr
- g_\k^{21}(r,r';\o)  i\si_r \pi_{\k}
& g_\k^{22}(r,r';\o) \pi_{-\k}		\cr} \right)\,,		
\ee
\eq{5defV}
V(\vr,\vrp;\o) = \sum_\k
\left( \matrix{ v_\k^{11}(r,r';\o) \pi_{\k}
& v_\k^{12}(r,r';\o) i\si_r \pi_{-\k} 	\cr
- v_\k^{21}(r,r';\o)  i\si_r \pi_{\k}
& v_\k^{22}(r,r';\o) \pi_{-\k}		\cr} \right) \,,
\ee
where the respective angular part is given by the $2\times 2$
matrices
\eq{5pi}
\pi_{\k}(\O,\O') \equiv \sum_\mu \cy_{\k\mu} (\O)
\cy_{\k\mu}^{\dag}(\O')\,.
\ee
The following reduction relationship holds:
\eq{O2}
\int d\O_2 \pi_{\k} (\O_1,\O_2)
 \pi_{\kp} (\O_2,\O_3) =\delta_{\k\kp} \pi_{\k}(\O_1,\O_3) \,.
\ee
$\cy_{\k\mu} (\O)$ are the usual two component spinor spherical
harmonics. They are eigenstates of the operators
$J^2,\; L^2$, $J_z$, and $K = (J +  1/2) (-1)^{(J-L+1/2)}$
\cite{Wi}
\eqn
\cy_{\k\mu}(\O) = \sum_{m_l,m_s} <l_\k \; m_l, {1\over 2}\;
m_s|j_\k
\; \mu> Y_{l_\k m_l}(\O) \chi_{m_s}
\ee
and obey the orthonormality relation
\eq{O1}
\int d\O \cy_{\k\mu}^{\dag}(\O)
 \cy_{\kp\mu'} (\O) =\delta_{\k\kp} \delta_{\mu\mu'} \,.
\ee

The radial functions $g$ and $v$ have the following symmetry
properties
\eqn
g^{ij}_\k(r,r';\o) = g^{ji}_\k(r',r;\o) = g^{ij}_\k(r,r';\o^*)^* \,.		
\ee

Inserting Eqs. (\ref{5defG}) and (\ref{5defV}) in the Dirac
\eqr{5dirac2} for the quark propagator and using (\ref{O1}) 
and
(\ref{O2}) yields
\[
\left[ \left( \matrix{-\o &-{1/r} -{\partial /\partial r}
+ {\k /r}\cr  {1/r} +{\partial /\partial r}
+ {\k /r} &-\o \cr} \right)
+ \left( \matrix{ {v_\k}^{11} &{v_\k}^{12}\cr
{v_\k}^{21} &{v_\k}^{22}\cr} \right) \right]
\left( \matrix{ {g_\k}^{11} &{g_\k}^{12}\cr
{g_\k}^{21} &{g_\k}^{22}\cr} \right)
\]
\eq{dvg}
= {\delta (r-r'\,) \over rr'} \,,
\ee
where $vg$ denotes $\int r^2_2 dr_2 v(r,r_2;\o)  g(r_2,r';\o)$
for notational convenience.

Defining $\bar g(r,r';\o) = rr' g(r,r';\o)$ and $\bar v(r,r';\o)
= rr' v(r,r';\o)$ Eq. (\ref{dvg}) simplifies finally to
\eq{5gbar}
\left[ \left( \matrix{-\o &-{\partial /\partial r}
+ {\k /r}\cr  {\partial /\partial r}
+ {\k /r} &-\o \cr} \right)
+ \left( \matrix{ {\bar v_\k}^{11} &{\bar v_\k}^{12}\cr
{\bar v_\k}^{21} &{\bar v_\k}^{22}\cr} \right) \right]
\left( \matrix{ {\bar g_\k}^{11} &{\bar g_\k}^{12}\cr
{\bar g_\k}^{21} &{\bar g_\k}^{22}\cr} \right)
= {\delta (r-r'\,) } \,.				
\ee
Details of the non-trivial numerical solutions of this 
equation are discussed in Sec. \ref{sec:numerical}.


\subsection{Radial part of the Schwinger-Dyson equation}
\label{subsec:radial}

Inserting (\ref{5defG}) and (\ref{5defV}) into  \eqr{5sd2}, 
we can write
\eqa
&&V(\vr,\vrp; y) =
\left( \matrix{ v_\k^{11}(r,r';\o) \pi_{\k}
& v_\k^{12}(r,r';\o) i\si_r \pi_{-\k} 	\cr
- v_\k^{21}(r,r';\o)  i\si_r \pi_{\k}
& v_\k^{22}(r,r';\o) \pi_{-\k}		\cr} \right)
\label{5a} \\
& \cr
& = & \al'\int dy' D_{\mu\nu}(\vr,\vrp; y') \al^\mu
G(\vr,\vrp; y-y') \al^\nu 	\cr
& \cr
& = & -\al'\int dy' d_{jll'}(r,r'; y') \bsi \cdot \bcY_{jlm}
\left(  \matrix{ g_\k^{22}(r,r';\o) \pi_{-\k}	
&- g_\k^{21}(r,r';\o)  i\si_r \pi_{\k}\cr
 g_\k^{12}(r,r';\o) i\si_r \pi_{-\k} 	
 & g_\k^{11}(r,r';\o) \pi_{\k} \cr} \right)
\bcY_{jl'm}^{\, *} \cdot \bsi	\cr
& \cr
& + & \al'\int dy' d_{l}^0 (r,r'; y') Y_{lm}
\left( \matrix{ g_\k^{11}(r,r';\o) \pi_{\k}
& g_\k^{12}(r,r';\o) i\si_r \pi_{-\k} 	\cr
- g_\k^{21}(r,r';\o)  i\si_r \pi_{\k}
& g_\k^{22}(r,r';\o) \pi_{-\k}		\cr} \right)
Y_{l m}	^* \,.
\label{5v}
\ea
From the Appendix we find
\eq{5coeffa}
\sum_{m} \bsi \cdot \bcY_{jlm} (\O) \pi_{\k}(\O,\O')
\bsi \cdot \bcY_{jl'm}^{\, *} (\O') = \sum_{\k'}
\cA^{\k' \k}_{j l l'} \, \pi_{\kp}(\O,\O') \,,
\ee
\eq{5coeffb}
\sum_{m} \bsi \cdot \bcY_{jlm}(\O) \si_r \pi_{\k}(\O,\O')
\bsi \cdot \bcY_{jl'm}^{\, *}(\O')
= \sum_{\kp } \cB^{\k' \k}_{j l l'}
\, \si_r \pi_{\kp} (\O,\O') \,,
\ee
\eq{5coeffc}
\sum_{m } Y_{lm}(\O) \pi_{\k}(\O,\O') Y_{lm}^*(\O')
= \sum_{\kp} \cC^{\k' \k}_l \, \pi_{\kp } (\O,\O') \,.
\ee
The self-energy coefficients $\cA,\cB$ and $\cC$ are explicitly
defined in the Appendix.

With these formulae the radial Schwinger-Dyson \eqr{5v} reads
\eqa
\bar v_\k^{11}(r,r';y) &=& \al' \int dy' \biggl[ d_l^0 (r,r')
\bar g_\kp^{11} (r,r';y') \cC^{\k \kp}_l    \cr
&-& d_{jll' }(r,r';y')\bar g_\kp^{22} (r,r';y-y')
\cA^{\k -\kp}_{j l l'} \biggr] \,,
\label{5sd1}
\ea
\eqa
\bar v_\k^{12}(r,r';y) &=& \al' \int dy' \biggl[ d_l^0 (r,r')
\bar g_\kp^{12} (r,r';y') \cC^{-\k -\kp}_l   \cr
&+& d_{jll' }(r,r';y')\bar g_\kp^{21} (r,r';y-y')
\cB^{-\k \kp}_{j l l'} \biggr] \,,
\ea
\eqa
\bar v_\k^{21}(r,r';y) &=& \al' \int dy' \biggl[ d_l^0 (r,r')
\bar g_\kp^{21} (r,r';y') \cC^{\k \kp}_l   \cr
&+& d_{jll'}(r,r';y')\bar g_\kp^{12} (r,r';y-y')
\cB^{\k -\kp}_{j l l'} \biggr] \,,
\ea
\eqa
\bar v_\k^{22}(r,r';y) &=& \al' \int dy' \biggl[ d_l^0 (r,r')
\bar g_\kp^{22} (r,r';y') \cC^{-\k -\kp}_l   \cr
&-& d_{jll'}(r,r';y')\bar g_\kp^{11} (r,r';y-y')
\cA^{-\k \kp}_{j l l'} \biggr] \,.
\label{5sd4}
\ea

Here we have also used Eqs. (\ref{5D}) and (\ref{5D0}) as well
as the symmetry properties $d^0_{l}(r,r') = d^0_{l}(r',r)$ and
$d_{jll'}(r,r';\o) = d_{jl'l}(r',r;\o)$ which hold for both pure
real and imaginary $\o$ because in Eq. (\ref{4tran}) for the
gluon propagator only $\o^2$ (and not $\o$) appears.


\section {THE QUARK WAVE FUNCTION}
\label{sec:wave}

By interpreting the nonlocal self-energy as an effective
potential we can now determine the wave function $q (\vr )$ and
energy eigenvalue $\ep$ of a single quark in the cavity. The
corresponding Dirac equation reads
\eq{5dir}
\bal \cdot {\b p} \ q (\vr ) + \int d^3 r_2 V(\vr,\vr_2;\ep) q
(\vr_2 ) = \ep \ q (\vr ) \,.
\ee
In spherical coordinates, $q (\vr )$ can be written in the form
\cite{Wi}
\eqn
q (\vr ) = \sum_{\k\mu} \left( \matrix{ u_{\k} (r )/r \cr
-{\rm i} \si_r  v_{\k} (r) /r \cr}
\right) \otimes \cy_{\k \mu} (\O) \,.
\ee
After angular decomposition, the radial part of \eqr{5dir}
obeys
\eqn
\left( \matrix{0 & - {\partial / \partial r}
+ {\k /r}\cr  {\partial / \partial r}
+ {\k /r} & 0 \cr} \right) \left( \matrix{  u_\k (r) \cr v_\k
(r)\cr} \right)
+ \int dr_2  {\bar V_\k}(r,r_2;\ep)
  \left( \matrix{  u_\k (r_2) \cr  v_\k (r_2)\cr} \right)
= \ep \left( \matrix{  u_\k (r) \cr  v_\k (r)\cr} \right) \,.
\ee


\section{NUMERICAL CALCULATION}
\label{sec:numerical}

In our calculations we use a (modified) Fermi function shaped
spherically symmetric color-dielectric function
\eq{5kappa}
\k (r) = { 1 - \k_v \over 1 + e^{(r-R)/A} } + \k_v  \,,	
\ee
where $R$ and $A$ are the radius and the surface thickness
respectively of the profile (see fig. 1). The small
but non-zero vacuum
value $\k_v$ guarantees that, e.g., the energy of a single quark
in the cavity remains finite. For color-singlet multi-quark
systems the limit $\k_v \rightarrow 0$ can be performed as will
be shown in Sec. \ref{sec:hadronic}.

Since our model is not renormalizable, an ultraviolet momentum
cutoff
is needed; this is consistent with asymptotic freedom.
This cutoff should reflect the energy scale of the
described physics; we choose $\Lambda_{CDM} = 5.0$ fm$^{-1}$.
In
terms of the variables used in this paper the $\o'$-integration
in
Eq. (\ref{5sd}) is cutoff at
$|\o_{max}| = \Lambda_{CDM}$ and the necessary summations over
angular momenta are limited by $l_{max} = R \ \o_{max}$ with $R$
from
Eq. (\ref{5kappa}). A careful analysis of the renormalization
problem for a nonlocal, spatially varying dielectric medium can
be found in ref. \cite{RWRG}.

With the gluon propagator derived in Sec. \ref{sec:gluon}, we
solve the coupled equations (\ref{5gbar}) and (\ref{5sd1} -
\ref{5sd4}) to obtain the full quark propagator and the quark
self-energy
on the imaginary $\o$-axis. Because of the absence of poles in
this region the self energy is numerically stable and no
oscillations
occur. A Taylor-expansion method is subsequently applied to
construct the quark self-energy on the real $\o$-axis:
\eqn
v_\k(r,r';z) = v_\k(r,r';0) + z v'_\k(r,r';0)
 + {z^2 \over 2} v^{\prime\prime}_\k(r,r';0)
 + {z^3 \over 6} v^{(3)}_\k(r,r';0)
 + {z^4 \over 24} v^{(4)}_\k(r,r';0)
 + \cdots \;,				
\ee
where the derivatives are evaluated in terms of the discrete
values
of the functions along the imaginary $\o$-axis.

\eqr{5gbar} is a coupled system of integro-differential
equations.
For its numerical solution we use matrix inversion.
It is well known that the leap frog
instability \cite{leap} (p. 342) appears in an equation like
\eqr{5gbar}
when the first order derivative is replaced by a centered
difference.
Therefore we introduce a small second order derivative term
to suppress the instability:
\[
\left[ \left( \matrix{-\o &-{\partial /\partial r}
+ B {\partial^2 /\partial r^2}
+ {\k /r}\cr  {\partial /\partial r}
+ B {\partial^2 /\partial r^2}
+ {\k /r} &-\o \cr} \right)
+ \left( \matrix{ {\bar v_\k}^{11} &{\bar v_\k}^{12}\cr
{\bar v_\k}^{21} &{\bar v_\k}^{22}\cr} \right) \right]
\]
\eqn
\times \left( \matrix{ {\bar g_{B\k} }^{11} &{\bar g_{B\k}
}^{12}\cr
{\bar g_{B\k} }^{21} &{\bar g_{B\k} }^{22}\cr} \right)
 = \delta (r-r'\,) \,,			\label{5gB}
\ee
where $B = b \ \Delta$ is a small number. $\Delta$ is the
grid interval and $b\sim \pm 1$ for $\kappa = \mp 1$ (the sign
is important to suppress the leap frog effect!). This additional 
regularizing term does not spoil the accuracy of the
solution.

Numerically we find that $g_\k = g_{B\k}$ satisfies \eqr{5gbar}
very well if $B\sim \pm\Delta$. Therefore $g_{B\k}$ can be
considered
as a first approximation to $g_{\k}$. There
must be a discontinuity in the Green's function for first order
differential equations. In \eqr{5gbar}, this discontinuity
occurs in
its off diagonal elements \cite{Gy}. We find numerically that
the additional second derivative term smooths ot the discontinuity
somewhat.

This approximation can be improved. This will be demonstrated
first in general terms. Consider the following two Green's
equations:
\eqn
L_0 G_0 = \delta \,,
\ee
\eq{4l}
(L_0 + L_B)G = \delta \,.
\ee
After operating with $G_0$ on both sides of \eqr{4l} and
integrating, we have
\eqa
G_0 &=&  G_0(L_0 + L_B )G \cr
&=&  G + G_0 L_B  G\,,
\ea
or
\eqn
G = G_0 - G_0 L_B G\,.			
\ee
Similarly, by integrating both sides of \eqr{5gbar} with $\bar
g_{B\k}$, we have the exact relation:
\eqa
&&\bar g_\k(r,r';\o) = \bar g_{B\k}(r,r';\o) \cr
&& + \int dr_2 \bar g_{0\k}(r,r_2;\o)
\left( \matrix{0 & B {\partial^2 /\partial r^2} \cr
B {\partial^2 /\partial r^2} &0 \cr} \right)
\bar v_\k(r_3,r_2;\o)  \bar g_\k (r_2,r';\o)\,.	
\label{5correct}	
\ea
This equation can be solved by iteration. However, in this case
the
leap frog instability eventually creeps in again. We have thus carried out
only one iteration.

Some of the results of the self energy calculation is shown in
figs. 4, 5 and 6. The figures display $\bar
V_\k^{mn}(\o;r,r')$ for $\k = -1$, $\o = 1 $fm$^{-1}$ and $(mn)
= (11), (22)$ and $(12)$ respectively.

We note that the self energy is non-zero. This
implies a dynamical breaking of chiral symmetry. The
structure of the self energy reflects the nonlocal character of
the
interaction. However, the self energy is sharply peaked around
$r = r'$ reflecting the dominance of the local contribution.

The self energy is inserted in the Dirac equation which is
solved self-consistently for the ground state ($\k = -1$). The
result is shown in fig. 7 for the $\k (r)$ profile
of fig. 1.

The single-quark energy $\ep$ is shown in fig. 8
as a function of $\k_v$. It does not exhibit a sign of
divergence as far as the calculation could be carried out (down
to $\k_v=0.05$). In fact, $\ep$ turns out to be quite
insensitive to $\k_v$ if $\k_v$ is small. The presented results
are thus gratifying.

We have tested our numerical calculation by varying the
following numerical parameters: 

(a) the number $N_{max}$ of $r$ grid points, 

(b) the integral limit $r_{max}$ of $r$ and 

(c) the number $N_\o$ of $\o$ grid points. 

For the actual parameters we have chosen the physical observables 
are all insensitive to them.


\section{HADRONIC PROPERTIES}
\label{sec:hadronic}

Having calculated the wave function and energy eigenvalue of a
single quark in a cavity, we now investigate color-neutral
composite systems of $N_q$ valence quarks. Evidently, $N_q = 2$
for mesons and $N_q = 3$ for baryons. The energy of these
systems is calculated in
the one gluon exchange approximation. Finally, corrections due
to the
center-of-mass motion and to the $\si$-field are taken into
account approximately.
Using scaling relations we fix the mass of the nucleon $m_N$
and
study the pion mass $m_\pi$. Its deviation from zero is a
measure of
how good our approximations are since the pion should be
massless
according to Goldstone's theorem.


\subsection{One gluon exchange approximation}
\label{subsec:oge}

The one gluon exchange interaction energy between quarks
(of equal eigen-energy) is given by \cite{Wi}
\eqn
E_{ex} = \al' \ \int d^3 r_1 d^3 r_2 [j^0(\vr_1) D^{00}(\vr_1,
\vr_2)
j^0(\vr_2) - \vj(\vr_1) \cdot \tensor D (\vr_1, \vr_2;0)
\cdot\vj(\vr_2) ] \,,	
\ee
with $\al' = {1 \over 4} g_s^2 \sum_{i < j} \br\pmb\lambda_i
\cdot
\pmb\lambda_j\ke$. The color matrix element $\br\pmb\lambda_1
\cdot
\pmb\lambda_2\ke$ has the value $-16/3$ for the pion and $-8/3$
for
the nucleon \cite{Wi}. Taking into account that
that for both the pion and the nucleon, the quarks are in the
ground state with $\k=-1, \mu = \pm 1/2$ the corresponding
currents can be evaluated and the exchange energy is ready 
calculated.

The total energy of quarks and gluons in a hadron with $N_q$
valence quarks is then given by
\eqn
E_{q,g} = N_q \ \ep + E_{ex} \,.
\ee


\subsection{Corrections and sigma contributions}
\label{subsec:corrections}

Up to now the $\si$-field has been neglected. However, it
contributes
to the total energy of the bag. The $\si$-field can be
reconstructed from $\k(r)$ and $\k(\si )$ given in Eqs.
(\ref{5kappa}) and (\ref{2kappa}) respectively. Then the
$\si$-field
energy is given by
\eqn
E_\si = \int \left[ {1\over 2} (\nabla \si)^2
+ U(\si)\right]	d^3r \,,
\ee
with the potential $U(\si)$ given in Eq. (\ref{usig}).
The total energy of the bag is then $E_{bag} = E_{q,g} + 
E_{\si}$.

We now address the hadronic center-of-mass energy.
Since  localization of the bag breaks Lorentz
invariance,  the bag acquires a non-zero total momentum that
contributes to the total energy of the system. The easiest way
to
correct this effect is to use the following approximate formula 
(projection \cite{LUB} would be better but more cumbersome):
\eqn
m^2_h = E_{bag}^2 - <P^2>_{bag}\,,	\label{mcorr}	
\ee
with
\eqn
<P^2>_{bag} = N_q \ <P^2>_q + <P^2>_\si \,.		
\ee
The momentum squared of one quark is given by
\eqn
<P^2>_q = \int d^3r |\nabla q|^2 \,.			
\ee
In order to calculate $<P^2>_\si$, the coherent state
approximation
\cite{Wi} is used:
\eqn
<P^2>_\si = \int d^3k k^2 \o_k f_k^2 \,.		
\ee
$f_k$ are the Fourier transforms of $\si(r)$, $\o_k$ is the
$\si$-field energy in the mode $k$.

For slowly varying $\o_k$ we finally get:
\eqn
<P^2>_\si = (m_{GB}^2 + <k^2>)^{1\over 2} \int d^3r (\nabla
\si)^2  \,,		
\ee
with the glueball mass $m_{GB}$.


\subsection {Scaling and the nucleon mass}
\label{subsec:scaling}

Scaling can be used to generate new solutions\cite{GW} from
those
presented so far. The equations are invariant under scale
transformations where all
lengths $r$ are replaced by
\be
r\ \to\ r' = \l r \,,
\ee
all energies and frequencies (including the cutoff
$\Lambda_{CDM}$) are replaced by
\be 
E \to\ E' = E/\l
\ee
and
\be
a\ \to\ a' = a/\l^2\,,\quad b\,'\ \to b/\l\,.
\ee
$c$ and $\a_s$ are invariant. The $\si$-field and the gluon
field potentials scale as length$^{-1}$.


\subsection {Numerical results}
\label{subsec:results}

Throughout our calculations, we use a cutoff $\Lambda_{CDM} = \o_m = 5 \ 
$fm$^{-1}$. With $l_m = R \ \o_m$, the quark wave functions and energies 
depend on the two parameters $R$ and $A$ from the $\k (r)$ profile. The 
hadron masses additionally depend on the parameters $a, b, c$ of the 
$\si$-field potential $U (\si )$. In order to minimize the number of free 
parameters, we assume that $U (\si )$ is universal in all hadrons. 
However, each hadron has a different $\k (r)$ profile reflecting the fact 
that the hadronic size is not universal.

The  numerical procedure is as follows: We choose a
potential $U (\si )$ and calculate the corrected nucleon mass
according to Eq. (\ref{mcorr}). Using scaling relations we
renormalize all dimensional properties by fixing the nucleon
mass to its empirical value $m_N = 938$ \ MeV. With these renormalized 
properties we now calculate the pion mass as a function of $R$ and $A$.

To this point, the $\si$-field is not self-consistent. We 
now
vary the parameters of the $\k (r)$ profile in order to find
an
extremum in the energy. This is a first approximation to a
fully self-consistent treatment. However, we expect the results
to be
reasonable since the proper shape of the $\si$-field is similar
to a Fermi function.

We find that there is not always a minimum in $A$ for a given
$m_\pi
(R)$. This may be due to the crude method used to correct the
effects
of the-center-of mass motion, to the form of $\kappa(r)$, or the
point is
an extremum, not a minimum.
The resulting pion mass is small but non-zero.


\section{SUMMARY AND PROSPECTIVES}
\label{sec:summary}

Within the framework of the chirally-invariant
chromo-dielectric
soliton model, the Abelian gluon propagator is solved in
configuration space for
a color-dielectric function with two parameters. The quark self
energy was obtained by solving the (nonlocal) Schwinger-Dyson
equation in configuration space as a function of imaginary
energy.
Quark wave functions and real eigenvalues were obtained. Bag
states
were constructed for the pion and the nucleon including one
gluon
exchange mutual interaction between quark pairs. The parameters
of
the parameterized $\si$-field (or equivalently, the dielectric
function $\k (r)$) were varied to extremize the bag energy.
Approximate center-of-mass corrections are calculated.
Employing scaling relations, the nucleon mass was set to its
empirical value. The resulting pion mass was determined to be
small
(the actual value depending on the model parameters) but not
zero, as demanded by Goldstone's theorem.

Extensions of the present work include the following:

a) Center-of-mass corrections based on variation after
projection.
This technique has been studied extensively by L\"ubeck {\it et
al.}
\cite{LUB} for the Friedberg-Lee soliton and was found to give
significant
corrections. It is certainly more reliable than the prescription 
$m^2
\approx <H>^2 - <p^2>$ used in the present paper.

b) A ``more'' self-consistent treatment of the soliton field by
either
solving the differential equation for the $\si$-field or by
including
more parameters in the functional form of $\k (\si )$ and $\k
(r)$.

c) Calculation of the mutual gluon exchange between quark pairs
by full summation of ladder diagrams.

d)  A systematic adjustment of model-parameters to fit the
properties
of all low-lying hadrons. This is not as tedious a task as it
might first
appear. The parameters of the model are $a,\ b,\ c$ and $\a_s$.
The
functional form of $\k( \si )$ also introduces a model
dependence,
but the results appear to be quite insensitive to that. Of the
four,
one is set by the nucleon mass using scaling from any given
set.
Results appear to be relatively insensitive to the ``family"
characterized by $b^2/ac$ \cite{Wi} but this is related to the
glueball mass which is assumed to lie in the range of $1 - 2 \ 
GeV$.
The key parameters include the nucleon size, magnetic moments,
$g_A /
g_V$ and the $N$-$\D$ mass splitting. Other hadronic spectra
properties are then regarded as predictions of the model.


\begin{center}
{\bf ACKNOWLEDGMENTS}
\end{center}

We are grateful to W. K\"opf, G. Krein and A. Williams for extensive 
discussions during all phases of this work. S.~H. 
wishes to thank the German Academic Exchange Service (DAAD) and the 
Cusanuswerk for financial support. This work was supported 
in part by the U.~S. Department of Energy.


\begin{appendix}
\section{}

In this appendix the self-energy coefficients $\cA,\cB$ and
$\cC$ from Sec.
\ref{subsec:radial} (Eqs. (\ref{5coeffa}-\ref{5coeffc})) are
explicitly evaluated.

We start with formula $(5.9.15)$ of Edmonds' \cite{Ed}
\eqn
\sigma_q\chi_\nu = \sqrt{3} <1/2,\,
\nu,\,1,\,q|1/2,\,q+\nu>\chi_{q+ \nu} \,.
\ee
Note that we use throughout our calculations the phase
convention of Edmonds. With the definitions
\eqn
\cy_{\k\mu}(\O)\equiv \cy_{j_\k \mu}^{l_\k}(\O)
\equiv \sum_{\nu m} <l_\k, m, 1/2, \nu|j_\k,\mu> Y_{l_\k m} 
\chi_\nu \,, \label{Ydef}
\ee
\eqn
\bcY_{ll'm}(\O) \equiv \sum_{q m'} <l' , m', 1 , q | l, m> Y_{l'
 m'} \vep_q \,,	
\ee
\eqn
\vep_{\pm 1} = \mp { \hat {\b x} \pm {\rm i} \hat {\b y} \over 
\sqrt 2 },\quad
\vep_0 = \hat {\b z} \,,				
\ee
we get
\eqa
&&\bcY_{ll'm} (\O)\cdot \bsi \cy_{\k\mu}(\O)	\cr
&=&\sum_{m_1 q m_2 \nu} < l',\, m_1,\, 1,\, q| l,\, m> Y_{l'm_1}
\si_{q}
< l_\k,\, m_2,\, 1/2,\, \nu| j_\k,\, \mu> \chi_\nu Y_{l_\k m_2} \cr
&=& \sum_{ m_1 q m_2 \nu LM\nu' }
< l',\, m_1,\, 1,\, q| l,\, m> Y_{L M}
< l_\k,\, m_2,\, 1/2,\, \nu| j_\k,\, \mu> < l',\, 0,\, l_\k,\,
0| L,\, 0> \cr
&\times &< l',\, m_1,\, l_\k,\, m_2| L,\, M>
\sqrt{ {(2l'+1)(2l_\k+1)}\over {4\pi(2L+1)} }
\sqrt{3 } < 1/2,\, \nu,\, 1,\, q| 1/2,\, \nu'>
\chi_{\nu'} \cr
&=&\sum_{\mu' j L} (-1)
\sqrt{ (2l+1)(2j_\k+1)(2l_\k+1)(2l'+1)3/2\pi} \cr
&\times &<l, \, m,\, j_\k,\, \mu| j,\, \mu'> < l',\, 0,\, l_\k,
 0| L, 0>
\left\{\matrix{ L& 1/2 &j \cr l'&1 &l \cr l_\k &1/2 &j_\k
\cr}\right\} \cy_{j\mu'}^L \,.
\label{Avec}			
\ea
Here we have used the contraction formula for spherical
harmonics (Edmonds $(5.16)$)  and the definition of the
$9j$-symbols (Edmonds $(6.4.3)$).

Similarly,
\eqa
&&2\hat {\b r} \cdot \bcY_{ll'm} (\O)
\cy_{\k\mu }(\O) \cr
&=& \sum_{m_1 q m_2 \nu} 2< l', m_1,\, 1, q| l, m> Y_{l'm_1}
\sqrt{4\pi \over 3} Y_{1q} < l_\k,\, m_2,\, 1/2,\, \nu| j_\k,\,
\mu> \chi_\nu Y_{l_\k m_2}	\cr
&=& \sum_{ m_1 q m_2 \nu } 2< l',\, 0,\, 1, 0| l, 0> Y_{lm}
 \sqrt{ {4\pi \over 3}\, {(2+1)(2l'+1)\over 4\pi (2l+1)} }
< l_\k,\, m_2,\, 1/2,\, \nu| j_\k,\, \mu> \chi_\nu Y_{l_\k m_2}
\cr
&=&\sum_{jL\mu'} (-1)^{1/2+l+l_\k+j} \sqrt{ 
(2l_\k+1)(2j_\k+1)(2l'+1)\over \pi}
< l',\, 0,\, 1,\, 0| l,\, 0>  \cr
&\times& < l,0,l_\k,0 |L,0>
\left\{\matrix{ j_\k  &1/2 &l_\k \cr L &l &j \cr}\right\}
< l,\, m,\, j_\k,\, \mu| j,\, \mu'> \cy_{j\mu'}^L
\label{Bvec} 
\ea
and
\eqa
&&Y_{lm} (\O)\cy_{\k\mu }(\O)			\cr
&=& \sum_{m_1 q m_2 \nu} Y_{lm}
< l_\k,\, m_2,\, 1/2,\, \nu| j_\k,\, \mu> \chi_\nu Y_{l_\k m_2}
\cr
&=&\sum_{jL\mu'} (-1)^{1/2-j_\k+l_\k+2j}
\sqrt{ (2l_\k+1)(2j_\k+1) (2l+1)\over 4\pi}		\cr
&\times & <l,0, l_\k ,0 |L ,0> \left\{\matrix{ j_\k &1/2 &l_\k
\cr L &l &j \cr}\right\} <  j_\k,\, \mu,\,l,\, m| j,\, \mu'>
\cy_{j\mu'}^L \,.		
\ea

According to \eqr{Avec}, the expression $\sum_{m\mu} \bcY_{ll'm}
(\O) \cdot \bsi \cy_{\k\mu }(\O) \cy_{\k\mu}^{ \dag} (\O') \bsi
\cdot \bcY_{l\lpp m}^* (\O')$
is proportional to $\delta_{jj'}$ and $\delta_{\mu\mu'}$. Now
$L$ and $L'$ have to  be equal or differ by $1$. However, $<
l',\, 0,\, l_\k,\, 0| L,\, 0> < \lpp , 0,\, l_\k, 0| L', 0>$
vanishes if $\vert L - L' \vert$ is odd, since $l' - \lpp$ is 
even, so only terms with $L = L'$ (or $\kp = \kpp$) contribute.
Thus
\eqa
&&\sum_{m\mu} \bcY_{ll'm} (\O) \cdot \bsi \cy_{\k\mu }(\O)
\cy_{\k\mu}^{ \dag} (\O') \bsi \cdot \bcY_{l\lpp m}^* (\O') 
\cr
&=& \sum_{\k \mu'} \cA^{\k' \k}_{l l' \lpp} \,
\cy_{\kp\mu'}(\O)
\cy_{\kp\mu' }^{\dag}  (\O') \,. 
\label{aaa}					
\ea
The following symmetry relation $\cA^{\k' \k}_{l l' \lpp} =
\cA^{\k' \k}_{l \lpp l'}$ holds.

Similarly, according to Eq. (\ref{Ydef}), the expression
$\sum_{m\mu}2\hat{\b r} \cdot \bcY_{ll'm}(\O) \cy_{\k\mu }(\O)
\cy_{\k\mu }^{\dag} (\O') \bsi \cdot \bcY_{l\lpp m}^* (\O')$ is 
proportional to $\delta_{jj'}$ and $\delta_{\mu\mu'}$. Now $L$ 
and $L'$ have again to be equal or differ by $1$. However, $< l,
 0,\, l_\k, 0| L, 0>
< l', 0,\, 1,0 | l, 0> < \lpp , 0,\, l_\k, 0| L', 0>$  vanishes 
if $\vert L - L' \vert$ is even, since $l' - \lpp$ is even, 
so only terms with $L = L' \pm 1$ (or $\kp = -\kpp$) contribute.
Thus
\eqa
&&\sum_{m\mu} 2\hat{\b r} \cdot \bcY_{l\lpp m}(\O)
\cy_{\k\mu }(\O) \cy_{\k\mu }^{\dag} (\O')
\bsi \cdot \bcY_{l\lpp m}^* (\O')
\cr
&=& \sum_{\k \mu'} \tilde \cB^{\k' \k}_{l l' \lpp} \, 
\cy_{\bar\kp\mu'} (\O)
\cy_{\kp\mu' }^{\dag} (\O') \,.
\label{bbb}
\ea

Similarly, we have
\eqn
\sum_{m\mu} Y_{lm} \cy_{\k\mu }(\O)
\cy_{\k\mu }^{\dag} (\O') Y_{lm}^*(\O')
= \sum_{\k \mu'} \cC^{\k' \k}_l \, \cy_{\kp\mu'} (\O)
\cy_{\kp\mu' }^{\dag} (\O')				
\ee

Finally, the quark-gluon coupling coefficients $\cA$, $\cB$ and
$\cC$ are given by
\eqa
\cA^{\k' \k}_{l l' \lpp}
&=&{3 \over 2\pi} \sqrt{ (2l'+1)(2\lpp
+1)}(2j_\k+1)(2l_\k+1)(2l+1)
< l', 0,\, l_\k, 0| l_\kp, 0>   \cr
&\times& < \lpp , 0,\, l_\k, 0| l_\kp, 0> \left\{\matrix{
l_\kp&1/2 &j_\kp\cr
l'& 1 &l \cr l_\k& 1/2 &j_\k \cr}\right\} \left\{\matrix{ l_\kp&
1/2 &j_\kp\cr
\lpp& 1  &l \cr l_\k& 1/2 &j_\k \cr}\right\} \,,	
\label{AGQ}	
\ea
\vfill
\eqa
&& \tilde \cB^{\k' \k}_{l l' \lpp}  \cr
&=& (-1)^{-1/2 +l+l_\k + j_{\kp}}
\sqrt{ 3/2 (2l'+1)(2\lpp +1)(2l+1)}{(2l_\k+1) (2j_\k+1) \over
\pi} 	\cr
&\times& < \lpp , 0,\, l_\k, 0| l_\kp, 0 >
< l', 0,\, 1, 0| l, 0> <l,0,l_\k ,0| l_{\bar\kp},0 >  \cr
&\times& \left\{\matrix{ j_\k &1/2  &l_\k   \cr
l_{\bar\kp} &l &j_{ \kp} \cr}\right\}
\left\{\matrix{ l_\kp &1/2 &j_\kp\cr
\lpp &1 &l \cr l_\k &1/2 &j_\k \cr}\right\}  \,, 
\ea
\eqn
\cC^{\k' \k}_l = { (2l_\k+1)(2j_\k+1) (2l+1)\over 4\pi} <l,0 
,l_\k ,0|l_\kp,0>^2
\left\{\matrix{ j_\k &1/2 &l_\k \cr l_\kp &l &j_\kp
\cr}\right\}^2 \,.
\ee
Furthermore,
\eqa
&&\sum_{m\mu} \bcY_{ll'm} (\O) \cdot \bsi \si_r \cy_{\k\mu 
}(\O)
\cy_{\k\mu }^{\dag} (\O') \bsi \cdot \bcY_{l\lpp m}^*(\O') 
\cr
&=&\sum_{m\mu} [2\hat{\b r} \cdot \bcY_{l\lpp m} (\O) -
\si_r \bcY_{ll'm} (\O) \cdot \bsi ]
\cy_{\k\mu }(\O) \cy_{\k\mu }^{\dag} (\O')
\bsi \cdot \bcY_{l\lpp m}^* (\O') 
\cr
&=& \sum_{\k \mu'} \cB^{\k' \k}_{l l' \lpp}
\, \si_r \cy_{\kp\mu'} (\O) \cy_{\kp\mu'} ^{\dag} (\O') \,.
\ea
In that very last step we have used Eqs. (\ref{Avec}) and
(\ref{Bvec}) as well as the identity $\si_r \ \cy_{\k\mu} = - 
\cy_{\bar\k\mu}$ and the identification
\eqn
\cB^{\k'\k}_{l l' \lpp}  \equiv - \cA^{\k' \k}_{l l' \lpp}
- \tilde \cB^{\k' \k}_{l l' \lpp} \,.				
\ee
Working out the hermitian conjugate of Eqs. (\ref{aaa}) and 
(\ref{bbb}) we get the following symmetry relations: $\cA^{\k' 
\k}_{l l' \lpp} = \cA^{\k' \k}_{l \lpp l'}$ and $\cB^{\k' \k}_{l
l' \lpp} =  \cB^{-\k' \ -\k}_{l \lpp l'}$.

\end{appendix}





\newpage

\begin{center}
{\Large Figures}
\end{center}

\bigskip

Figure 1: 
The color-dielectric function $\k (r)$ for $R = .8$ fm, $A = .15$ fm and 
$\k_v = .15$ ($r$ in fm). 

\medskip



Figure 2:
The tensor part of the gluon propagator in the transverse magnetic mode 
$d_{102}(r,r')$.

\medskip

Figure 3:
The tensor part of the gluon propagator in the transverse magnetic mode 
$d_{122}(r,r')$.

\medskip

Figure 4:
The quark self energy on the real $\o$-axis: 
$\bar v_{-1}^{11}(r,r')$.

\medskip

Figure 5:
The quark self energy on the real $\o$-axis: $\bar
v_{-1}^{22}(r,r')$.

\medskip

Figure 6:
The quark self energy on the real $\o$-axis: 
$\bar v_{-1}^{12}(r,r')$.

\medskip

Figure 7:
The quark wave function. $r u(r)$ is the darker line,
$r v(r)$ is the lighter one.

\medskip

Figure 8:
The single-quark energy $\ep$ (in fm$^{-1}$) as a function of $\k_v$



\newpage

\begin{table}
\centering
\caption {Table of pion masses for
$a=39.9$,
$b=        -746.2$,
$c=        4569.6$,
$B=            0.03892$,
$m_{GB}=       1310.8$}

\begin{tabular}{||r|r|r|r|r|r|r||}
$R$  &$A$ &$E_q$       &$E_{q,g}$      &$\sqrt{<P^2>}_Q $
&$\sqrt{<P^2>}_\si$  &$m_\pi$ \\
(fm)    &(fm)      &(MeV)     &(MeV)     &(MeV)     &(MeV)&(MeV)
\\
\hline
0.6  &0.150     &404.92    &610.61    &477.20    &189.6
&171.31\\
0.6  &0.175     &405.81    &612.17    &476.50    &178.3
&235.94\\
0.6  &0.200     &405.97    &613.19    &476.10    &170.1
&290.23\\
0.6  &0.225     &405.71    &615.40    &475.90    &164.0
&344.06\\
0.6  &0.250     &405.22    &614.81    &475.80    &159.4
&390.66\\
\hline
0.8  &0.150     &295.37    &452.89    &489.10    &249.4
&344.64\\
0.8  &0.175     &296.77    &454.19    &486.50    &232.1
&284.70\\
0.8  &0.200     &297.20    &454.17    &484.80    &219.0
&208.37\\
0.8  &0.225     &297.01    &453.40    &483.70    &208.8
&61.61\\
0.8  &0.250     &296.39    &452.22    &482.90    &200.7
&198.74\\
\hline
1.0  &0.150     &230.69    &358.91    &485.20    &311.7
&338.67\\
1.0  &0.175     &232.25    &360.33    &480.00    &288.5
&238.99\\
1.0  &0.200     &232.76    &360.09    &476.20    &270.6
&27.71\\

1.0  &0.225     &232.60    &358.83    &473.80    &256.3 
&237.19\\
1.0  &0.250     &231.99    &356.95    &471.70    &244.8
&342.74\\
\hline
1.2  &0.150     &194.97    &308.29    &400.90    &375.5
&371.10\\
1.2  &0.175     &196.45    &309.63    &398.10    &346.6
&462.86\\
1.2  &0.200     &197.08    &309.56    &396.10    &323.9
&540.69\\
1.2  &0.225     &197.18    &308.61    &394.60    &305.7
&612.00\\
\end{tabular}
\end{table}

\end{document}